\documentstyle[preprint,aps,epsf]{revtex}
\draft
\begin{document}
\preprint{Nucl. Phys. A (1997) submitted.}
\title{Chemical and mechanical instability in hot isospin-asymmetric 
nuclear matter}
\bigskip
\author{\bf Bao-An Li\footnote{email: Bali@comp.tamu.edu} 
and C.M. Ko\footnote{email: Ko@comp.tamu.edu}}
\address{Cyclotron Institute and Department of Physics\\ 
Texas A\&M University, College Station, TX 77843, USA}
\maketitle

\begin{quote}
Using phenomenological equations of state, we study the chemical and 
mechanical instabilities in hot isospin-asymmetric nuclear matter.
Both instabilities are found to depend strongly on the isospin asymmetry 
of the nuclear matter. For a chemically and mechanically stable asymmetric 
nuclear matter, a lower neutron excess is found in the liquid phase than 
in the gas phase. Furthermore, the neutron excess in the mixed phase at 
densities between that of the liquid and the gas phases is limited by
certain maximum value.  Also, for the nuclear equations of state considered
here the boundary of chemical instability in the pressure-density plane is 
found to be more extended than that of mechanical instability. However, 
locations of these boundaries depend sensitively on both the bulk 
compressibility of nuclear matter and the density dependence of the symmetry 
energy.
\end{quote}

\newpage
\section{Introduction}
During the last decade, there have been extensive studies on 
multifragmentation in nuclear reactions, which is characterized by the 
emission of several intermediate mass fragments with a power law distribution.
It has been suggested that this phenomenon is due to various mechanical 
instabilities of the nuclear matter, such as the volume instability, the 
surface instability of the Rayleigh kind, and the Coulomb instability. A 
recent review on nuclear multifragmentation can be found in ref. \cite{moretto}.  
With radioactive ion beams, which have been successfully developed recently, 
nuclear matter with a high isospin asymmetry can be created transiently 
during reactions induced by neutron-rich nuclei (e.g., 
\cite{tanihata95,hansen95}). This thus offers the possibility to study also 
the chemical instability associated with the isospin asymmetry of the 
nuclear matter. Indeed, recent experiments on nuclear reactions involving 
different ratios of the neutron to proton numbers have shown that 
products from nuclear multifragmentation depends strongly on the isospin 
asymmetry of the colliding nuclei \cite{sherry,kunde96,dem96}.
   
It is well-known that nuclear matter is not thermodynamically stable at all 
densities $\rho$, temperatures $T$, and neutron excess $\delta\equiv 
(\rho_n-\rho_p)/\rho$. The necessary and sufficient conditions for 
the existence of a stable isospin-asymmetric nuclear matter can be expressed 
by the following inequalities (e.g., \cite{lat78,bar80,muller}):
\begin{eqnarray}
\left(\frac{\partial E}{\partial T}\right)_{\rho,\delta}&>& 0,\\\
\left(\frac{\partial P}{\partial \rho}\right)_{T,\delta}&\geq& 0,\\\
\left(\frac{\partial \mu_n}{\partial \delta}\right)_{P,T}&\geq& 0,\label{chem}
\end{eqnarray}
where $E$, $P$, and $\mu_n$ are the energy per nucleon, pressure and 
neutron chemical potential, respectively. The first condition is required by
the thermodynamic stability and is always satisfied by any reasonable nuclear 
equation of state; the second condition ensures the mechanical stability 
against the growth of density fluctuations; and the last one protects the 
chemical or diffusive stability (DS) against the development of 
neutronization. The last two conditions can be violated in some regions 
of the $(\rho, T, \delta)$ configuration space, leading to mechanical 
and/or chemical instabilities.

The boundaries of mechanical and chemical instabilities in the 
$(\rho, T, \delta)$ space depend on the equation of state (EOS) of 
asymmetric matter. In the Relativistic Mean-Field (RMF) theory, which has a 
linear density dependence in the symmetry energy, recent studies indicate 
that the diffusive spinodal associated with the chemical instability of 
asymmetric nuclear matter encloses more of the configuration space than 
the isothermal spinodal associated with the mechanical instability 
\cite{muller}. For values of isospin asymmetry, $\delta\leq 0.4$, 
which are reachable in reactions induced by neutron-rich nuclei, the 
separation between the isothermal and diffusive spinodals in $P-\rho$ 
plane is significant, and the diffusive instability becomes a more 
relevant spinodal in discussing the properties of asymmetric nuclear matter. 

However, the density dependence of nuclear symmetry energy is poorly 
known, and theoretical predictions vary widely 
\cite{siemens70,baym71,laga81,wiringa88,prak88,lat91,thor94}.
Although it is linear in density in the RMF theory, a $\rho^{1/3}$ dependence 
was obtained by Siemens using the Bethe-Goldstone theory for asymmetric 
nuclear matter \cite{siemens70}.  More sophisticated calculations by 
Wiringa {\it et al.} using the variational many-body theory \cite{wiringa88} 
give different density dependence depending on the nuclear forces used 
in the calculations. It is thus both interesting and necessary to examine 
how the structure of mechanical and chemical instabilities in asymmetry 
nuclear matter might change when other nuclear equations of state are used.  

The nuclear symmetry energy at densities other than the normal nuclear
matter density is of interest in both astrophysics and nuclear physics. 
For example, the cooling rate, chemical composition and magnetic moment 
of a neutron star depend critically on this \cite{lat91,sumi95,kut93}.  
For nuclear reactions induced by neutron-rich nuclei, it was recently 
found that the nuclear symmetry energy affects significantly the reaction 
dynamics, especially the ratio of preequilibrium neutrons to protons 
emitted in the early stage of heavy-ion collisions at intermediate 
energies \cite{likr}. 

In the present paper, we study the chemical and mechanical instabilities of 
asymmetric nuclear matter using various phenomenological equations of state. 
These EOS's have been used previously by us in studying the ratio of 
preequilibrium neutrons to protons in nuclear reactions induced by 
neutron-rich nuclei. In particular, we examine the dependence of the 
boundaries of these instabilities on both the compressibility and the 
density dependence of the symmetry energy. We find that for all equations 
of state considered here a similar conclusion as that based on the RMF 
is obtained, i.e., the isothermal spinodal is enclosed within the diffusive 
spinodal, and the location of these two spinodals depends sensitively on 
both the bulk compressibility of nuclear matter and the density dependence 
of the symmetry energy. 

\section{Thermodynamical analysis of chemical and mechanical instabilities}
For asymmetric nuclear matter in thermal equilibrium, the density $\rho_q$ 
of neutrons ($q=n$) or protons ($q=p$) is
\begin{equation}
\rho_q=\frac{1}{\pi^2}\int_{0}^{\infty}k^2f_q(k)dk,
\end{equation}
where
\begin{equation}
f_q(k)=[{\rm exp}(e_q-\mu_q)/T+1]^{-1}
\end{equation}
is the Fermi distribution function, and $e_q$ is the single particle kinetic 
energy.  For modestly high temperatures ($T\geq 4 {\rm MeV}$), this equation 
can be inverted analytically to obtain the chemical potential 
\cite{brack,jaqaman1,jaqaman2}
\begin{equation}\label{muq}
\mu_q=V_q+T\left[{\rm ln}(\frac{\lambda_T^3\rho_q}{2})
+\sum_{n=1}^{\infty}\frac{n+1}{n}b_n(\frac{\lambda_T^{3}\rho_q}{2})^n\right],
\end{equation}
where 
\begin{equation}
\lambda_T=\left(\frac{2\pi \hbar^2}{mT}\right)^{1/2}
\end{equation}
is the thermal wavelength of the nucleon, and $b_n's$ are the inversion 
coefficients given in refs. \cite{brack,jaqaman1,jaqaman2}. In eq. 
(\ref{muq}), $V_q$ is the single particle potential energy and can be
parameterized as
\begin{equation}\label{pot}
      V_{q}(\rho,\delta) = a (\rho/\rho_{NM}) + b (\rho/\rho_{NM})^{\sigma}\ 
	+V_{{\rm asy}}^{q}(\rho,\delta).
\end{equation}
The parameters $a,~b$ and $\sigma$ are determined by the saturation 
properties and the compressibility $K$ of symmetric nuclear matter, 
i.e., 
\begin{eqnarray}
a&=&-29.81-46.90\frac{K+44.73}{K-166.32}~({\rm MeV}),\\
b&=&23.45\frac{K+255.78}{K-166.32}~({\rm MeV}),\\
\sigma&=&\frac{K+44.73}{211.05}.
\end{eqnarray}

The isospin-independent term should also contain a momentum-dependent 
part which is important for some dynamical observables, such as the 
collective flow (e.g., \cite{gale87,gale90,zhang}), but is not essential 
for other observables, such as the ratio of preequilibrium neutrons to 
protons \cite{likr}.  Effects of momentum-dependent interactions on thermal 
properties were studied recently in refs. \cite{csernai92,fai} and were
found to have only small effects.  We thus neglect the momentum-dependent 
interaction and concentrate on investigating effects of the 
isospin-asymmetric potential $V_{{\rm asy}}^{q}$. This potential is given by
\begin{equation}
V^{q}_{{\rm asy}}(\rho,\delta)=\partial w_a(\rho,\delta)/\partial \rho_{q},
\end{equation} 
where $w_a(\rho,\delta)$ is the contribution of nuclear interactions to the 
symmetry energy density, i.e.,  
\begin{equation}
w_a(\rho,\delta)=e_a\cdot \rho F(u)\delta^2,
\end{equation}
and
\begin{equation}
e_a\equiv S_0-(2^{2/3}-1)\frac{3}{5}E_F^0.
\end{equation}
In the above, $u=\rho/\rho_0$ with $\rho_0$ the normal nuclear matter 
density; $S_0$ is the symmetry energy $S(\rho)$ at $\rho_0$ and is known 
to be in the range of 27-36 MeV \cite{Hau88}; and $E_F^0$ is the Fermi 
energy at $\rho_0$. 

For the function $F(u)$, we consider the following three forms: $F(u)=u^2$, 
$u$ and $u^{1/2}$, which are taken form the results of typical 
microscopic many-body calculations. Then the corresponding symmetry potentials 
are, respectively, 
\begin{eqnarray}\label{vasy}
V_{{\rm asy}}^{n(p)}&=&\pm 2e_a u^2\delta+e_a u^2\delta^2,\\\
V_{{\rm asy}}^{n(p)}&=&\pm 2 e_a u\delta,
\end{eqnarray}
and 
\begin{equation}
V_{{\rm asy}}^{n(p)}=\pm 2e_a u^{1/2}\delta-\frac{1}{2}e_au^{1/2}\delta^2,
\end{equation}
where ``$+$'' and ``$-$'' are for neutrons and protons, respectively.

From the chemical potentials determined by eq. (\ref{muq}), we obtain the 
total pressure of the system from the Gibbs-Duhem relation,
\begin{equation}\label{gibbs}
\frac{\partial P}{\partial \rho}=\frac{\rho}{2}\left[(1+\delta)
\frac{\partial \mu_n}{\partial \rho}+(1-\delta)\frac{\partial 
\mu_p}{\partial \rho}\right].
\end{equation}
The result can be separated into three parts, i.e., 
\begin{equation}
P=P_{{\rm kin}}+P_0+P_{{\rm asy}},
\end{equation}
where $P_{{\rm kin}}$ is the kinetic contribution
\begin{equation}
P_{{\rm kin}}=T\rho\{1+\frac{1}{2}\sum_{n=1}^{\infty}
b_n(\frac{\lambda_T^{3}\rho}{4})^n\left[(1+\delta)^{n+1}
+(1-\delta)^{1+n}\right]\},
\end{equation}
and $P_0$ is the contribution from the isospin-independent nuclear 
interaction
\begin{equation}
P_0=\frac{1}{2}a\rho_0(\frac{\rho}{\rho_0})^2+\frac{b\sigma}
{1+\sigma}\rho_0(\frac{\rho}{\rho_0})^{\sigma+1}.
\end{equation}
$P_{{\rm asy}}$ is the contribution from isospin-dependent part of 
the nuclear interaction, and it can be written as 
\begin{eqnarray}
P_{{\rm asy}}&=&2e_a\rho_0(\frac{\rho}{\rho_0})^3\delta^2,\\\
P_{{\rm asy}}&=&e_a\rho_0(\frac{\rho}{\rho_0})^2\delta^2,
\end{eqnarray}
and
\begin{equation}
P_{{\rm asy}}=\frac{1}{2}e_a\rho_0(\frac{\rho}{\rho_0})^{3/2}\delta^2
\end{equation}
for $F(u)=u^2,~u$ and $u^{1/2}$, respectively.

The condition for mechanical stability follows from eq.\ (\ref{gibbs}) 
and can thus be determined straightforwardly. To evaluate the condition for
chemical stability, we use the following relations:
\begin{eqnarray}\label{trans}
\left(\frac{\partial \mu_n}{\partial \delta}\right)_{T,P}&=& 
\left(\frac{\partial \mu_n}{\partial \delta}\right)_{T,\rho} 
-\left(\frac{\partial \mu_n}{\partial P}\right)_{T,\delta} 
\cdot\left(\frac{\partial P}{\partial \delta}\right)_{T,\rho},\nonumber\\
&=&
\left(\frac{\partial \mu_n}{\partial \delta}\right)_{T,\rho} 
-\left(\frac{\partial \mu_n}{\partial \rho}\right)_{T,\delta} 
\cdot\left(\frac{\partial P}{\partial \rho}\right)^{-1}_{T,\delta}
\cdot\left(\frac{\partial P}{\partial \delta}\right)_{T,\rho}.
\end{eqnarray}

\begin{figure}[htp]
\setlength{\epsfxsize=14truecm}
\centerline{\epsffile{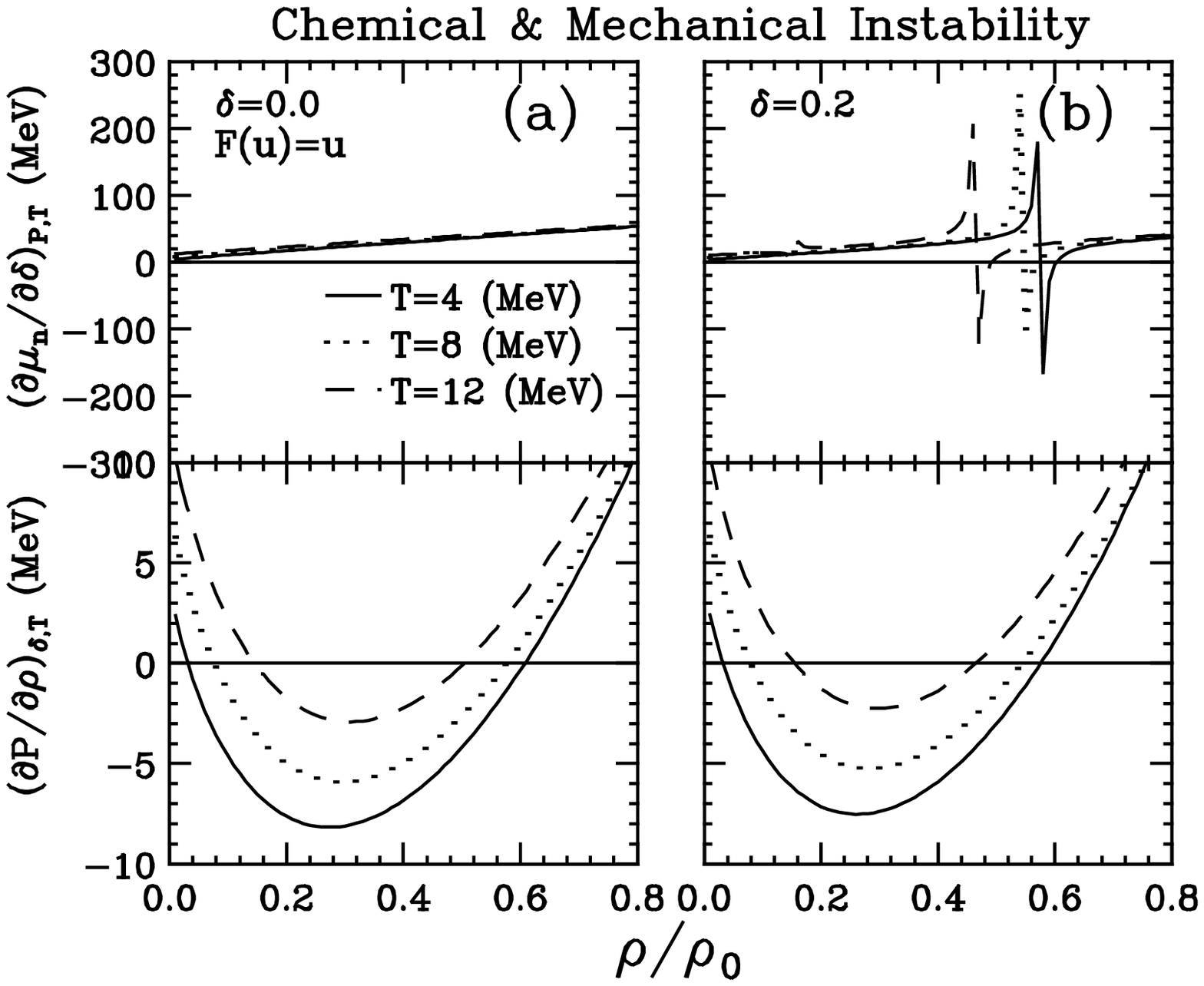}}
\caption{Chemical (upper window) and mechanical (lower window) stability 
conditions as functions of density at 
fixed temperatures $T=$ 4, 8 and 12 MeV for $\delta=0.0$ (left panel), 
and $\delta=0.2$ (right panel).}
\label{chem1}
\end{figure}

\begin{figure}[htp]
\setlength{\epsfxsize=14truecm}
\centerline{\epsffile{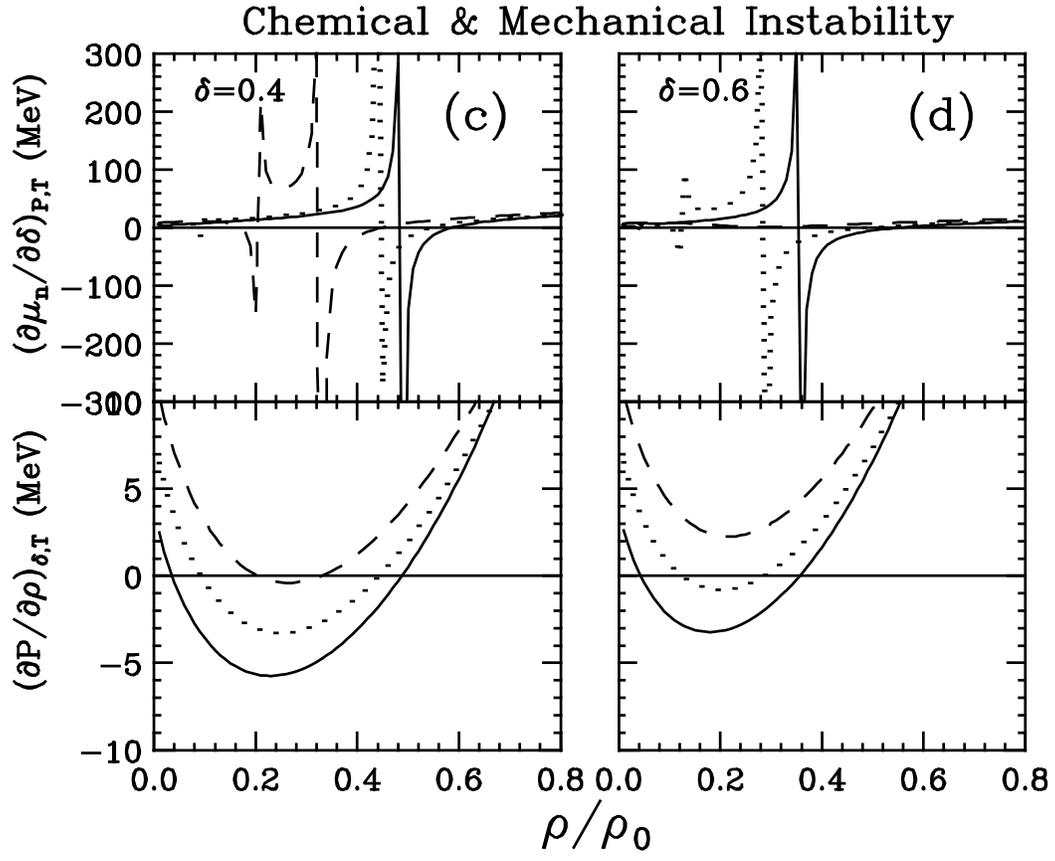}}
\caption{Same as Fig. \protect\ref{chem1} but for $\delta=0.4$ (left panel), 
and $\delta=0.6$ (right panel).}
\label{chem2}
\end{figure}

Knowing $\left(\frac{\partial \mu_n}{\partial \delta}\right)
_{T,P}$ and $\left(\frac{\partial P}{\partial \rho}\right)_{T,\delta}$ 
as functions of $\rho, T$ and $\delta$, we can then identify regions of 
the configuration space where chemical and/or mechanical instability 
occur. We first show in Fig. \ref{chem1} and Fig. \ref{chem2}
these two quantities as functions of $\rho$ at various $T$ and $\delta$.  
In the calculation $K=200$ MeV and $F(u)=u$ are used. For comparison we 
also show in the left panel of Fig. \ref{chem1} results for symmetric 
nuclear matter. In this case there is no chemical instability and the 
quantity $\left(\frac{\partial \mu_n}{\partial \delta}\right)_{T,P}$ 
increases with both density and temperature. To understand this, we 
note that
\begin{eqnarray} 
\left(\frac{\partial P}{\partial \delta}\right)_{T,\rho}
&=&\frac{2}{\delta}P_{{\rm asy}}+\frac{T\rho}{2}\sum_{n=1}^{\infty}(n+1)b_n
(\frac{\lambda_T^3\rho}{4})^n\left((1+\delta)^n-(1-\delta)^n\right),\\\
\left(\frac{\partial \mu_n}{\partial \delta}\right)_{T,\rho}
&=&\left(\frac{\partial V_{{\rm asy}}^n}{\partial \delta}\right)_{T,\rho}
+T\left[\frac{1}{1+\delta}+\sum_{n=1}^{\infty}(n+1)b_n
\left(\frac{\lambda_T^3\rho}{4}\right)^n(1+\delta)^{n+1}\right],
\end{eqnarray}
and 
\begin{equation}
 {\rm lim}_{\delta\rightarrow 0}
\left(\frac{\partial P}{\partial \delta}\right)_{T,\rho}=0.
\end{equation}
Then, it can be shown that
\begin{eqnarray}
&{\rm lim}&_{\delta\rightarrow 0} \left(\frac{\partial 
\mu_n}{\partial \delta}\right)_{T,P}= 
{\rm lim}_{\delta\rightarrow 0} \left(\frac{\partial \mu_n}{\partial \delta}
\right)_{T,\rho},\\\
&=&2e_aF(u)+T\left[1+\sum_{n=1}^{\infty}(n+1)b_n
\left(\frac{\lambda_T^3\rho}{4}\right)^n\right],
\end{eqnarray}
which is always positive as expected for symmetric nuclear matter.
Furthermore, it increases with both density and temperature.  

Mechanical instability is known to happen at intermediate densities 
corresponding to the mixed phase between the gas and liquid phases at 
subcritical temperatures. This is clearly demonstrated by plotting 
$\left(\frac{\partial P}{\partial \rho}\right)_{T,\delta}$ as a function of
$\rho$. It is interesting to examine this quantity at the low density limit.
From the Gibbs-Duhem relation of eq.\ (\ref{gibbs}) and the expression for
$\mu_q$ of eq.\ (\ref{muq}), we obtain
\begin{equation}
{\rm lim}_{\rho\rightarrow 0}\left(\frac{\partial P}
{\partial \rho}\right)_{T,\delta}
={\rm lim}_{\rho\rightarrow 0}(T+a\frac{\rho}{\rho_0})=T,
\end{equation}
as in a dilute gas.

\begin{figure}[htp]
\setlength{\epsfxsize=14truecm}
\centerline{\epsffile{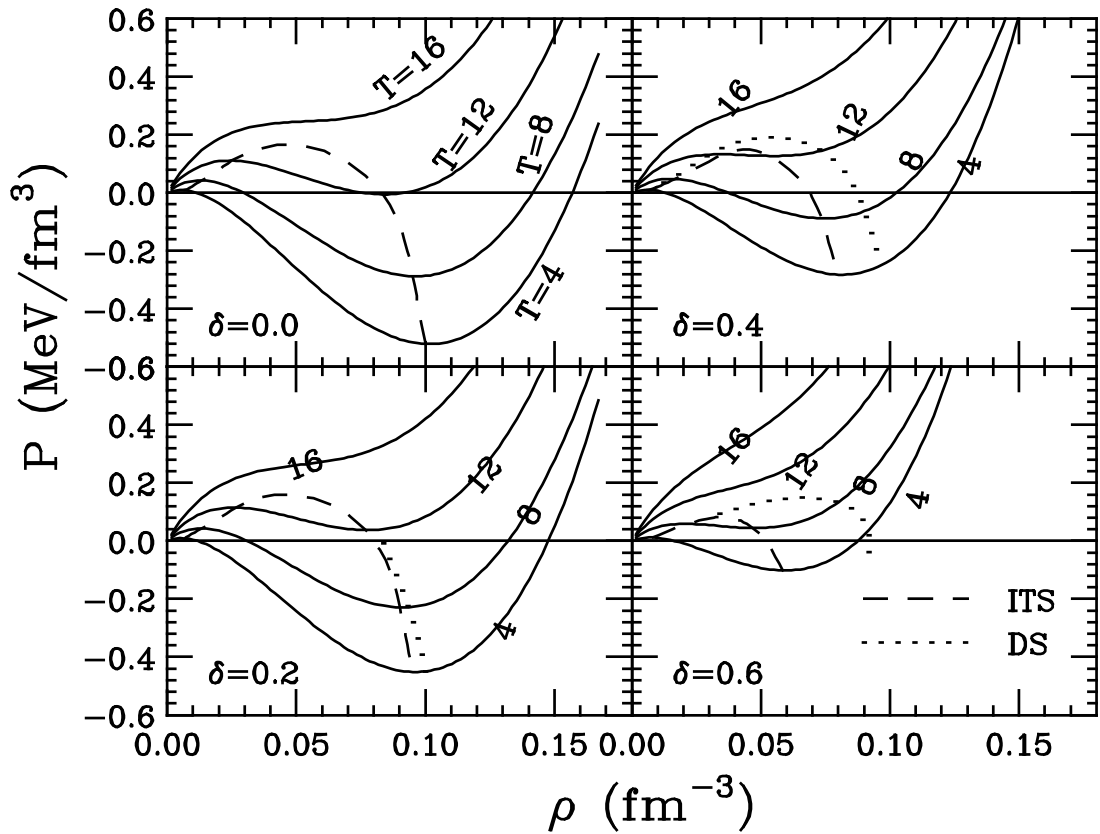}}
\caption{Pressure as a function of density at fixed temperatures $T=$
4, 8, 12 and 16 MeV for $\delta=$ 0.0, 0.2, 0.4 and 0.6. The isothermal 
(diffusive) spinodals are plotted using dashed (dotted) lines.}
\label{pre}
\end{figure}

We note from eq. (\ref{trans}) that the quantity $\left(\frac{\partial \mu_n}
{\partial \delta}\right)_{T,P}$ is singular along the boundary of mechanical 
instability where $\left(\frac{\partial P}{\partial \rho}\right)_{T,\delta}=0$.
This singularity is indicated by the spikes in Fig. \ref{chem1} and 
Fig. \ref{chem2}. However, the height of these spikes between the 
boundaries of the gas phase and the mixed phase is relatively low.  This is 
because at this boundary the last derivative in eq. (\ref{trans}) is also 
close to zero, i.e., 
\begin{equation}
{\rm lim}_{\rho\rightarrow 0}\left(\frac{\partial P}
{\partial \delta}\right)_{T,\rho}
=\frac{2}{\delta}\cdot {\rm lim}_{\rho\rightarrow 0}P_{{\rm asy}}=0.
\end{equation}
Although a nuclear system is always chemically stable in the low density
limit as
\begin{equation}
{\rm lim}_{\rho\rightarrow 0}\left(\frac{\partial \mu_n}{\partial \delta}
\right)_{P,T}=\frac{T}{1+\delta}>0,
\end{equation}
it starts to develop chemical instability at finite density once it 
becomes neutron-rich.  In particular, it is seen that the boundary of 
mechanical instability shrinks while that of chemical instability expands 
as the isospin asymmetry $\delta$ increases. For example, at a temperature 
of T=8 MeV the mechanical instability gradually disappears but the chemical 
instability becomes more pronounced as $\delta$ increases from 0 to 0.6. 
Furthermore, in the mechanically unstable region the nuclear system can 
be chemically stable, but at higher densities chemical instability happens
even in the mechanically stable region. This is more clearly shown in Fig. 
\ref{pre} where the pressure along the chemical (DS) and mechanical (ITS) 
spinodals are plotted as functions of density. Both the isothermal pressure 
and separation between the two spinodals increase with increasing $\delta$. 
Therefore, not only the chemical but also the mechanical instability is 
strongly isospin-dependent.  

\begin{figure}[htp]
\setlength{\epsfxsize=14truecm}
\centerline{\epsffile{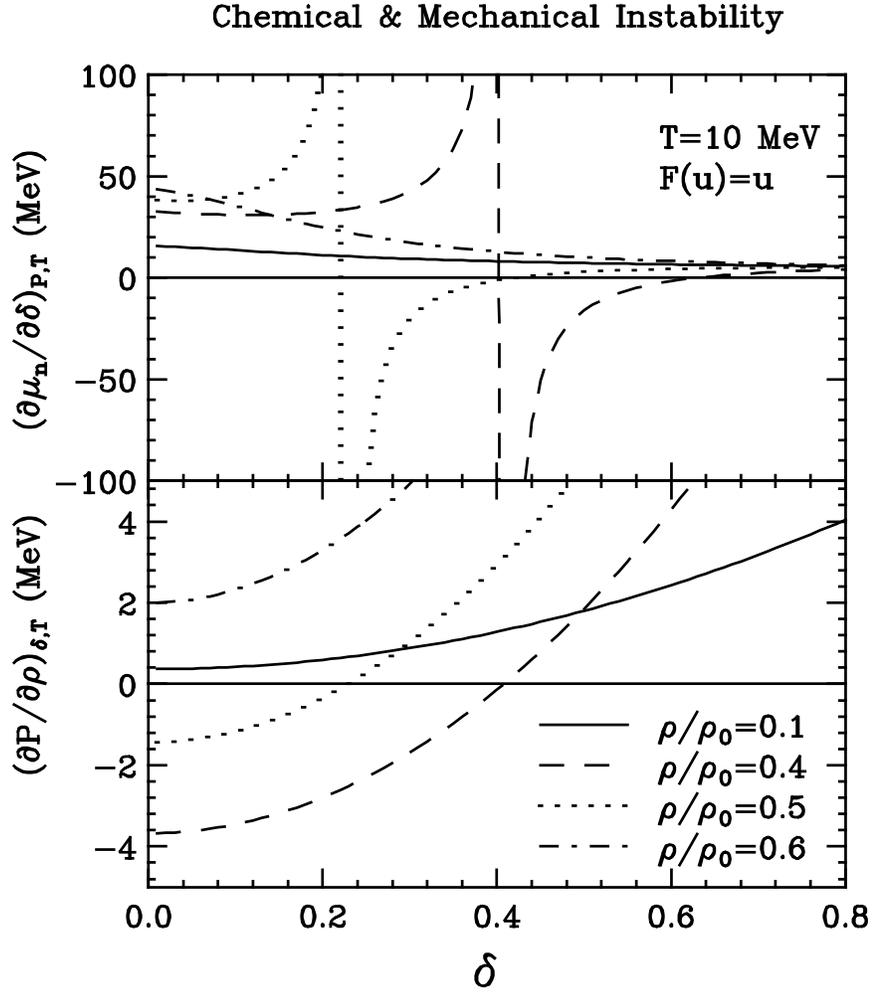}}
\caption{Chemical (upper window) and mechanical (lower window) stability 
conditions as functions of $\delta$ at a fixed temperature $T=10$ MeV and 
various densities.}
\label{chem3} 
\end{figure}

\begin{figure}[htp]
\setlength{\epsfxsize=14truecm}
\centerline{\epsffile{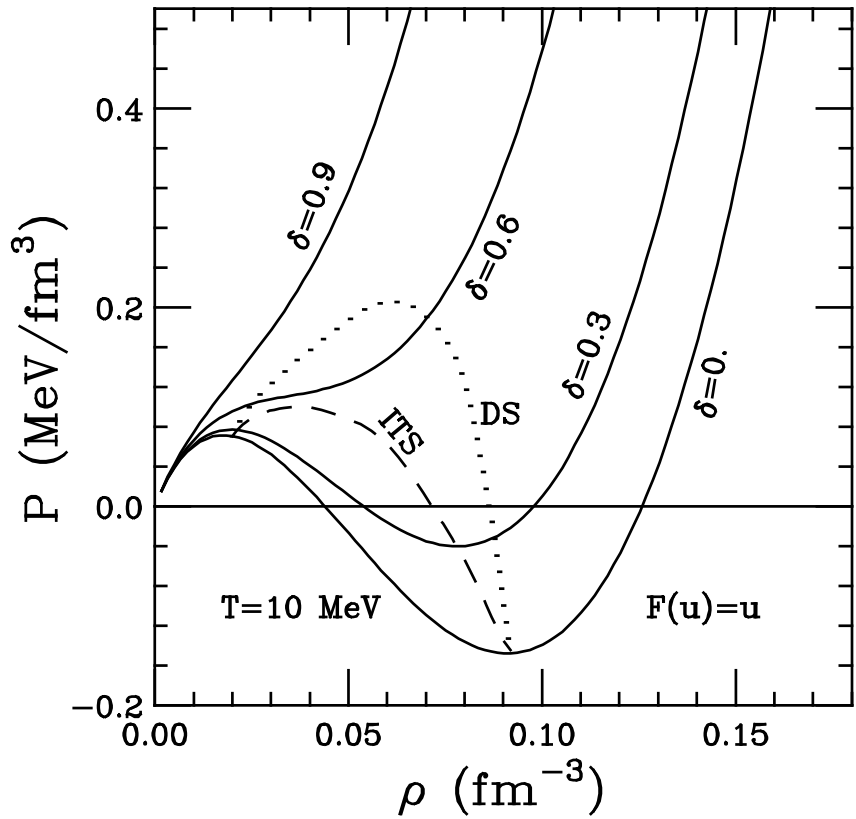}}
\caption{Pressure as a function of density at a
fixed temperature $T=10$ MeV at various $\delta$.
The isothermal (diffusive) spinodals are plotted using 
dashed (dotted) lines.}
\label{pre1}
\end{figure}

The isospin dependence of mechanical and chemical instabilities at a fixed 
temperature is studied in more detail in Fig. \ref{chem3} and Fig. 
\ref{pre1}. It is seen that the system is both mechanically and chemically 
stable at low and high densities.  At intermediate densities (e.g., 
$\rho/\rho_0=0.4, 0.5$) the system is mechanically unstable at low 
$\delta$ (e.g., $\delta\leq 0.2$ and $\delta\leq 0.4$ for $\rho/\rho_0=0.4$ 
and 0.5, respectively) and chemically unstable at intermediate $\delta$ 
(e.g., $0.2\leq \delta \leq 0.4$ for $\rho/\rho_0=0.4$ and $0.4\leq \delta 
\leq 0.6$ for $\rho/\rho_0=0.5$, respectively). The corresponding boundaries 
in the pressure-density plane are shown in Fig. \ref{pre1}. Again, the 
diffusive spinodal line is more extended than the isothermal spinodal. The 
isospin dependence of the pressure is also shown in this figure. We note 
that its strong dependence on $\delta$ is due to the significant 
$P_{{\rm asy}}$ contribution to the total pressure.

\begin{figure}[htp]
\setlength{\epsfxsize=14truecm}
\centerline{\epsffile{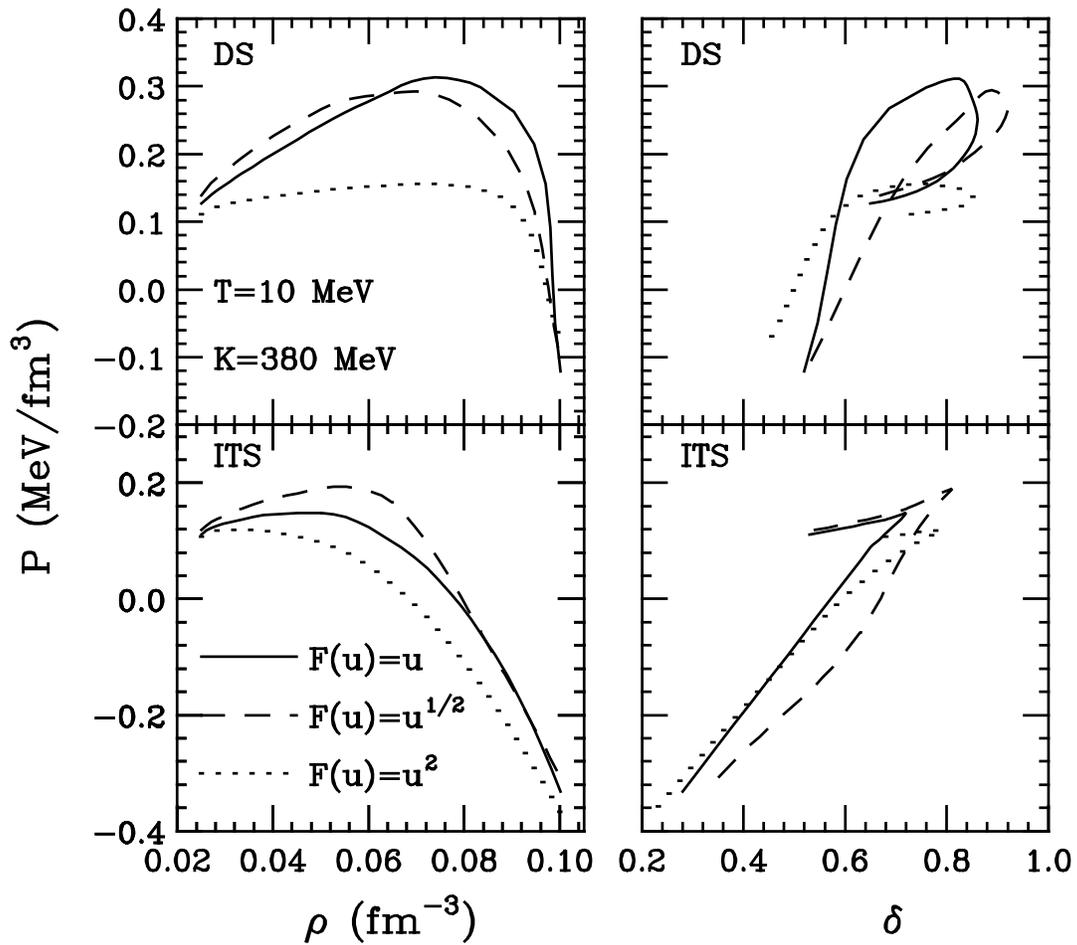}}
\caption{Pressure as a function of density (left panel) and isospin 
asymmetry (right panel) at a fixed temperature $T=10$ MeV along 
the boundary of diffusive spinodals (upper windows) and isothermal 
spinodals (lower windows) by using three forms of $F(u)$ and a 
compressibility of $K=380$ MeV.}
\label{k380}  
\end{figure}

\begin{figure}[htp]
\setlength{\epsfxsize=14truecm}
\centerline{\epsffile{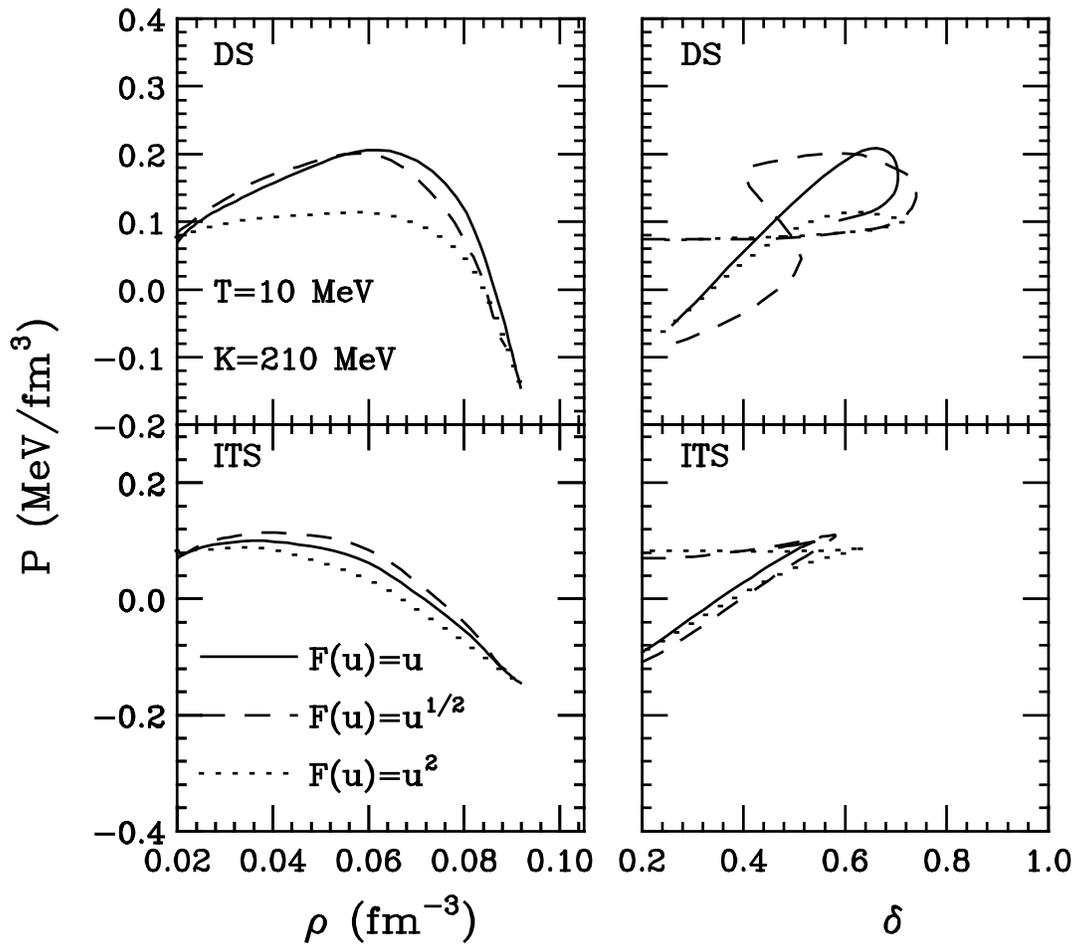}}
\caption{Same as Fig. \protect\ref{k380} for a compressibility of $K=200$ MeV.}
\label{k200}  
\end{figure}

Both chemical and mechanical instabilities also depend on the 
isospin-dependent and -independent parts of the nuclear equation of 
state.  To demonstrate this, we show in Fig. \ref{k380} and Fig. 
\ref{k200} the pressure as a function of density $\rho$ (left panels) 
and isospin asymmetry $\delta$ (right panels) along the diffusive spinodals 
(upper windows) and isothermal spinodals (lower windows) at a constant 
temperature of $T=10$ MeV by using the three forms of $F(u)$ and a 
compressibility of 380 and 200 MeV, respectively. From these two figures 
we observe the following interesting features. First, the diffusive spinodals 
are always more extended than the isothermal spinodals. Second, both DS and 
ITS depend on the form of $F(u)$, and this dependence is stronger for the 
stiff equation of state of $K=$ 380 MeV. Third, from the right panels of 
these two figures one sees that it is both chemically and mechanically 
favorable for the system to be less asymmetric (smaller $\delta$) in 
the liquid phase than in the gas phase. This result is consistent with
that based on the energy consideration. Since the equation of state
for asymmetric nuclear matter contains a $S(\rho)\delta^2$ term, 
it is therefore energetically favorable for it to separate into a liquid 
phase that is less asymmetric and a gas phase that is more asymmetric,
rather than into two phases with equal isospin asymmetry. Finally, 
it is seen that there is a maximum isospin asymmetry for both DS and ITS. 
Along both DS and ITS the isospin asymmetry increases to a maximum value 
and then decreases as the density is reduced from the liquid phase towards
the gas phase. As shown in both figures, the maximum isospin asymmetry 
is also sensitive to both the isospin-dependent and -independent parts of 
the nuclear equation of state.      

\section{Summary and discussions}
Using various phenomenological equations of state, we have studied
both the chemical and mechanical instabilities in hot isospin-asymmetric 
nuclear matter. Both instabilities are found to depend strongly on the 
isospin asymmetry of the nuclear matter. Also, in a chemically and mechanically 
stable asymmetric nuclear matter a lower neutron excess is favored in the 
liquid phase than in the gas phase.  Furthermore, there is a maximum neutron 
excess in the mixed phase at intermediate densities. For all nuclear equations
of state considered here, the boundary of chemical instability is found
to be more extended than that of mechanical instability. However, the 
location of these boundaries in the ($T$, $\rho$, $\delta$) space depends 
sensitively on both the bulk compressibility of nuclear matter and the 
density dependence of the symmetry energy. These results are relevant
in understanding the significant differences observed in the fragment
distributions from collisions involving isospin-symmetric and 
isospin-asymmetric nuclei at intermediate energies 
\cite{sherry,kunde96,dem96}.  One scenario for nuclear multifragmentation 
is that the hot system formed in a reaction expands almost adiabatically 
into the mechanical instability region and then disassemble into clusters 
and nucleons due to the growth of density fluctuations (e.g. \cite{bsiemens}).  
Great efforts have thus been devoted during the last decade (e.g., 
\cite{ayik1,randrup1,randrup2,colonna,ayik2}) to incorporate the effects of 
density fluctuations into dynamical transport models. To describe nuclear 
multifragmentation from heavy ion collisions involving asymmetric nuclei,
it will be of interest to extend these dynamical models to include also 
the isospin degree of freedom and its fluctuations.

\bigskip

This work was supported in part by the NSF Grant No. PHY-9509266. 
We would like to thank J.B. Natowitz for helpful discussions.

\newpage

\newpage

\newpage

\newpage

\newpage

\newpage

\newpage

\newpage

\end{document}